\documentclass[runningheads]{cl2emult}

\usepackage{makeidx}  % allows index generation
\usepackage{graphicx} % standard LaTeX graphics tool
\usepackage{subeqnar} % subnumbers individual equations
\usepackage{multicol} % used for the two-column index
\usepackage{cropmark} % cropmarks for pages without
\usepackage{lnp}      % placeholder for figures
\makeindex            % used for the subject index
\begin{document}
\title*{The extragalactic background and its fluctuations in the
far-infrared wavelengths}
\toctitle{The extra-Galactic background and its flucuations in the
far-infrared wavelengths}
% allows explicit linebreak for the table of content
%
%
\titlerunning{The extra-Galactic background and its fluctuations in the
far-infrared}
% allows abbreviation of title, if the full title is too long
% to fit in the running head
%
\author{G. Lagache\inst{1}
\and J-L. Puget\inst{1}
\and A. Abergel\inst{1}
\and F.R. Bouchet\inst{3} 
\and F. Boulanger\inst{1}
\and P. Ciliegi\inst{4}
\and D.L. Clements\inst{5} 
\and C.J. Cesarsky\inst{6} 
\and F.X. D\'esert\inst{2}
\and H. Dole\inst{1}
\and D. Elbaz\inst{6} 
\and A. Franceschini\inst{7}  
\and R. Gispert\inst{1} 
\and B. Guiderdoni\inst{3} 
\and L.M. Haffner\inst{8}
\and M. Harwit\inst{9} 
\and R. Laureijs\inst{10} 
\and D. Lemke\inst{11}  
\and A.F.M. Moorwood\inst{12} 
\and S. Oliver\inst{13}
\and W.T. Reach\inst{14} 
\and R.J. Reynolds\inst{8}
\and M. Rowan-Robinson\inst{13}
\and M. Stickel\inst{11}
\and S.L. Tufte\inst{15}}

\authorrunning{G. Lagache, J-L. Puget et al.}
% if there are more than two authors,
% please abbreviate author list for running head
%
%
\institute{Institut d'Astrophysique Spatiale, Orsay, France
\and Laboratoire d'Astrophysique, Observatoire de Grenoble, France
\and Institut d'Astrophysique de Paris, France
\and Osservatorio Astronomico di Bologna, Italy
\and Cardiff University, UK
\and Service d'Astrophysique, CEA/DSM/DAPNIA Saclay, France 
\and Osservatorio Astronomico di Padova, Italy
\and Astronomy Department, University of Wisconsin, Madison, USA
\and 511 H.Street S.W., Washington, DC 20024-2725
\and ISOC ESA, VILSPA, Madrid, Spain
\and MPIA, Heidelberg, Germany
\and ESO, Garching, Germany
\and Imperial College, London, UK
\and IPAC, Pasadena, CA, USA
\and Department of Physics, Lewis \& Clark College, Portland, USA}

\maketitle              % typesets the title of the contribution

\begin{abstract}
A Cosmic Far-InfraRed Background (CFIRB) has long been predicted
that would traces the intial phases of galaxy formation.
It has been first detected by~\cite{PUG96} using
COBE data and has been later confirmed by several recent studies 
(~\cite{FIX98},~\cite{HAU98},~\cite{LAG99}).
We will present a new determination of the CFIRB that uses for the first time, 
in addition to COBE data, two independent gas tracers: the HI survey
of Leiden/Dwingeloo (~\cite{HAR98}) and the WHAM H$_{\alpha}$ survey (~\cite{REY99}). We will see that
the CFIRB above 100 $\mu$m is now very well constrained.
The next step is to see if we can detect its fluctuations.
To search for the CFIRB fluctuations, we have used the FIRBACK observations. 
FIRBACK is a deep cosmological survey conducted at 170$\mu$m with ISOPHOT
(~\cite{DOL00}). We show that 
the emission of unresolved extra-galactic sources clearly dominates,
at arcminute scales, the background fluctuations
in the lowest galactic emission regions. This is the first
detection of the CFIRB fluctuations.
\end{abstract}

\section{Determination of the CFIRB above 100 $\mu$m: an other approach}
\begin{figure}
\centering
\includegraphics[width=.8\textwidth]{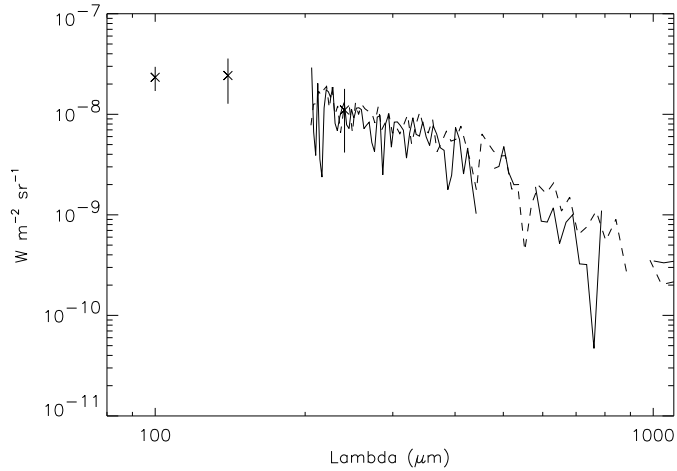}
\caption[]{CFIRB spectra obtained from the decomposition
of the far-infrared sky (continuous line) and determined for the Lockman
Hole region (dashed line) by~\cite{LAG99}. Also reported
are DIRBE values at 100, 140 and 240 $\mu$m.}
\label{CFIRB_spec}
\end{figure}

In very diffuse parts of the sky (no molecular clouds
and HII regions), the far-IR emission 
can be written as the sum of dust emission associated
with the neutral gas, dust associated with the ionised gas, 
interplanetary dust emission and the CFIRB (and eventually 
the cosmological dipole and
CMB). In previous studies (~\cite{PUG96},~\cite{FIX98},~\cite{HAU98}), 
dust emission associated with the ionised gas
which was totally unknown has been either not subtracted
properly or neglected. \\
We have detected for the first time
dust emission in the ionised gas (~\cite{LAG99})
and shown that the emissivity (which
is the IR emission normalised to unit hydrogen column density)
of dust in the ionised gas was nearly the same as that in the
neutral gas. This has consequences on the determination of the CFIRB.
Following this first detection, we have combined HI and 
WHAM H$_{\alpha}$ data (~\cite{REY99})
with far-IR COBE data in order to derive dust
properties in the diffuse ionised gas as well as to make a proper
determination of the CFIRB.
Technically, after a careful pixel selection
(see~\cite{LAG00} for more details) we describe the 
far-infrared dust emission as a function of the HI
and H$^+$ column density by:
\begin{equation}
\label{main_EQ}
IR= A \times N(HI)_{20cm^{-2}} + B \times N(H^+)_{20cm^{-2}} + C
\end{equation}
where N(HI)$_{20cm^{-2}}$ and N(H$^+$)$_{20cm^{-2}}$ 
are the column densities normalised
to 10$^{20}$ H cm$^{-2}$. The coefficients A, B
and the constant term C are determined simultaneously using regression fits. 
We show that about 25$\%$ of the IR
emission comes from dust associated with the ionised gas
which is in very good agreement with the first determination
(~\cite{LAG99}). 
The CFIRB spectrum obtained using this  
far-infrared emission decomposition is shown in Fig.~\ref{CFIRB_spec} 
together with the CFIRB FIRAS determination
of~\cite{LAG99} in the Lockman Hole region. We see a very
good agreement between the two spectra. These determinations are also
in good agreement with~\cite{FIX98}.

At 140 and 240 $\mu$m, the values obtained for the CFIRB
are 1.13$\pm$0.54 MJy/sr and 0.88$\pm$0.55 MJy/sr respectively 
For each selected pixel, we compute
the residual emission, R = IR - A$\times$N(HI) - B$\times$N(H$^+$).
Uncertainties of the CFIRB have been derived from the width of the histogram
of R (statistical uncertainties derived from the regression analysis are negligible). 
The obtained CFIRB values, although much more noisy
(due to the small fraction of the sky used), are in very
good agreement with the determination of ~\cite{HAU98}.
At 140~$\mu$m, the CFIRB value of~\cite{LAG99} is smaller 
than that derived here since the assumed WIM (Warm Ionised
gas) dust spectrum
was overestimated (the WIM dust spectrum was
very noisy below 200~$\mu$m and the estimated dust
temperature was too high).\\

At 100~$\mu$m, 
assuming an accurate subtraction of the zodiacal emission,
our decomposition gives:
I$_{CFIRB}$(100)= 0.78$\pm$0.21 MJy/sr. This is the first 
time that two independent gas tracers for the HI and the
H$^+$ have been used to determine the background at 100 
$\mu$m. One has to note that methods based on
the intercept of the far-IR/HI correlation
for the determination of the CFIRB 
are dangerous. For example, for our selected parts of the sky, 
this intercept is about 0.91 MJy/sr, which is quite different from
the value of the CFIRB (0.78 MJy/sr). 
The CFIRB value of 0.78 MJy/sr can be compared to the non-isotropic residual
emission found by~\cite{HAU98}. The average over three regions
of the residual emission, equal
to 0.73$\pm$0.20 MJy/sr, is in very good agreement with our
determination.\\

So we see, using different approaches, that we are now converging
on the shape and level of the CFIRB above 100 $\mu$m.
The next step is to see if we can detect its fluctuations
and study them.

\section{Why search for the CFIRB fluctuations?}
The CFIRB is made of sources with number counts as a function of flux which can
be represented, for the present discussion, by a simple power law:
$$N(>S)= N_0 \left( \frac{S}{S_0} \right)^{- \alpha}$$
Obviously, these number counts need to flatten at low fluxes
to insure a finite value of the background. Thus, we assume that $\alpha$=0 
for $S<S^{\ast}$.\\

For the simple Euclidian case ($\alpha$=1.5), the CFIRB integral 
is dominated by sources near S$^{\ast}$ and
its fluctuations are dominated by sources which are just below the detection
limit S$_0$. It is well known that strong cosmological
evolution, associated with a strong negative K-correction, could lead to a very steep
number count distribution (see for example~\cite{GUI98} and~\cite{FRA98}). 
In the far-IR present observations show a very steep slope of
$\alpha$=2.2 (~\cite{DOL00}). In this case,
the CFIRB integral is still dominated by sources near S$^{\ast}$ but
its fluctuations are now also dominated by sources close to 
S$^{\ast}$. Thus, it is essential to study the 
extra-galactic background fluctuations which are likely to be dominated
by sources with a flux comparable to those dominating the
CFIRB intensity.\\

To see if we can detect the CFIRB fluctuations, we need wide field
far-IR observations with high angular resolution and very high signal to noise ratio.
The FIRBACK project, which is a very deep cosmological survey
with ISOPHOT at 170 $\mu$m (~\cite{DOL00}), sustains all these conditions. 
To search for CFIRB fluctuations, we have first used the so-called
``Marano 1'' field for which we obtained 16 independent coadded
maps which allow us to determine very properly the instrumental noise.
In this field, we have a signal to noise ratio of about 300
and we detect 24 sources (~\cite{PUG99})
that we remove from the original map.
We then extend our first analysis to the other FIRBACK fields.
Details on the data reduction and calibration can be found
in ~\cite{LAG00b}.\\

Source subtracted maps show background fluctuations which are
made of two components that we want to separate, 
galactic cirrus fluctuations and if present the
extra-galactic ones. 

\section{Extra-galactic and galactic background fluctuation separation: detection of the CFIRB fluctuations}
\begin{figure}
\centering
\includegraphics[width=.8\textwidth]{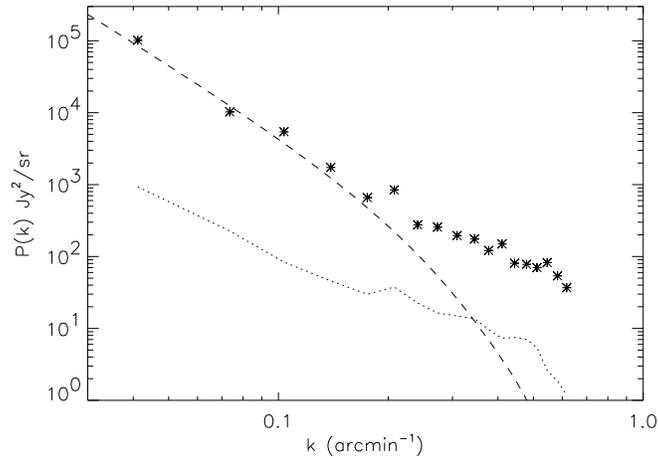}
\caption[]{Power spectrum of the source subtracted ``Marano 1'' field
($\ast$).
The instrumental noise power spectrum (dotted line) has been subtracted. The dashed line
represents the cirrus power spectrum, multiplied by the footprint.}
\label{fluc_Marano}
\end{figure}

Our separation of the extra-galactic and galactic fluctuations
is based on a power spectrum decomposition. This method
allows us to discriminate the two components using the statistical
properties of their spatial behaviour. Fig. ~\ref{fluc_Marano} 
shows the power spectrum of the ``Marano 1'' field.
In the plane of the detector, the power spectrum measured on the map 
can be expressed in the form:
\begin{equation}
P_{map}= P_{noise} + (P_{cirrus} + P_{sources}) \times W_k 
\label{Eq_Pk}
\end{equation}
where  P$_{noise}$ is the instrumental noise power spectrum measured
using the 16 independent maps of the Marano 1 field (~\cite{LAG00c}), 
P$_{cirrus}$ and P$_{sources}$ are the cirrus and
unresolved extra-galactic source power spectra respectively, 
and W$_k$ is the footprint power spectrum. For our analysis,
we remove P$_{noise}$ from P$_{map}$.\\

We know from previous work that the cirrus far-infrared 
emission power spectrum, P$_{cirrus}$,
has a steep slope in $\sim k^{-3}$ 
(~\cite{GAU92},~\cite{KOG96},~\cite{HER98},~\cite{WRI98},~\cite{SCH98}). 
These observations
cover the relevant spatial frequency range and have been recently extended
up to 1 arcmin using very diffuse HI data (~\cite{MAM99}). 
The extra-galactic
component is unknown but certainly much flatter (see the discussion in~\cite{LAG00c}). 
We thus conclude
that the steep spectrum observed in our data at k$<$0.15 arcmin$^{-1}$ 
(Fig.~\ref{fluc_Marano}) can only be due to cirrus emission. 
The break in the power spectrum at k$\sim$ 0.2 arcmin$^{-1}$ is 
very unlikely to be due to the cirrus emission itself which is 
known not to exhibit any prefered scale (~\cite{FAL98}). 
Thus, the normalisation of our cirrus power spectrum P$_{cirrus}$ is directly determined
using the low frequency data points and assuming a k$^{-3}$ dependence.\\

We clearly see in Fig.~\ref{fluc_Marano} an excess over P$_{cirrus}$ between
k=0.25 and 0.6 arcmin$^{-1}$ which is more than a factor of 10
at k=0.4 arcmin$^{-1}$. Any reasonable power law spectrum
for the cirrus component multiplied by the footprint
leads, as can be easily seen in Fig. ~\ref{fluc_Marano}, to a very steep
spectrum at spatial frequency k$>$0.2 arcmin$^{-1}$
which is very different from the observed spectrum.
Moreover, the excess is more than
10 times larger than the measured instrumental noise power spectrum.
Therefore, as no other major source of fluctuations
is expected at this wavelength, the large excess observed between
k=0.25 and 0.6 arcmin$^{-1}$ is interpreted as due
to unresolved extra-galactic sources.
This is the first detection of the CFIRB fluctuations.\\

The Marano 1 field cannot be used to
constrain the clustering of galaxies due to it rather small size.
However, the extra-galactic source power spectrum mean level can be determined.
We obtain P$_{sources}$~=~7400 Jy$^2$/sr, which is in very good
agreement with the one predicted by~\cite{GUI97}.
This gives CFIRB rms fluctuations around 0.07 MJy/sr (for
a range of spatial frequencies up to 5 arcmin$^{-1}$). 
These fluctuations are at the $\sim$9 percent level,
which is very close to the predictions of~\cite{HAI99}.\\

\begin{figure}
\centering
\includegraphics[width=.8\textwidth]{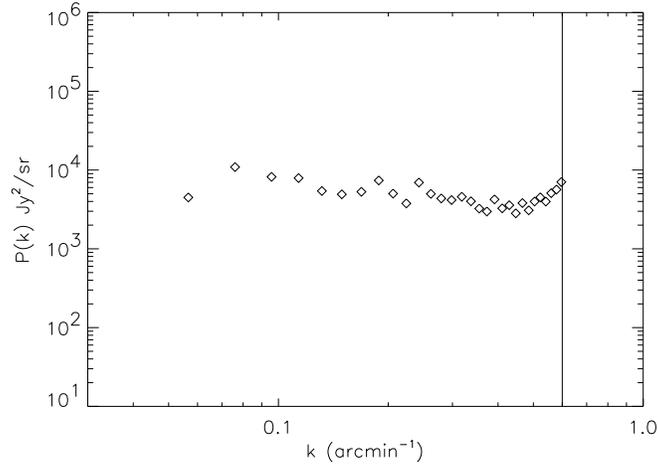}
\caption[]{Extra-galactic source power spectrum of the FIRBACK/ELAIS N2 field.
The vertical line shows the cut-off in angular resolution.}
\label{fluc_N2}
\end{figure}

The same analysis can be done for the other and larger FIRBACK
fields. From Eq. ~\ref{Eq_Pk}, we deduce: 
$$P_{sources}= (P_{map} - P_{noise}) / W_k - P_{cirrus}$$
Fig. ~\ref{fluc_N2} shows the extra-galactic fluctuation
power spectrum (P$_{sources}$) obtained for the FIRBACK/ELAIS N2 field. 
It is very well fitted with a constant CFIRB fluctuation power spectrum 
of about 5000 Jy$^2$/sr, which is in good agreement with that obtained
in the ``Marano 1'' field. We obtain also exactly the same
extra-galactic fluctuation power spectrum in the FIRBACK N1 field
with a value of about 5000 Jy$^2$/sr.

\section{Conclusions}
We have shown in the FIRBACK fields that the extra-galactic background
fluctuations lie well above the instrumental noise
and the cirrus confusion noise. The observed power spectrum shows a flattening
at high spatial frequencies which is due to unresolved extra-galactic 
sources.
The level of the extra-galactic power spectrum fluctuations
is nearly the same in all FIRBACK fields.
The next step consists of removing the cirrus contribution using independent 
gas tracers (the H$_{\alpha}$ and the 21cm emission lines)
to isolate the extra-galactic fluctuation
brightness and try to constrain the IR large scale structures.\\

%%%%%%%%%%%%%%%%%%%%%%%%%%%%%%%%%%%%%%%%%%%%%%%%%%%%%%%%%%%%%%%%%%%%%%%%%%%%%%%%%%

%INDEX%%%%%%%%%%%%%%%%%%%%%%%%%%%%%%%%%%%%%%%%%%%%%%%%%%%%%%%%%%%%%%%
\clearpage
\addcontentsline{toc}{section}{Index}
\flushbottom
\printindex
%%%%%%%%%%%%%%%%%%%%%%%%%%%%%%%%%%%%%%%%%%%%%%%%%%%%%%%%%%%%%%%%%%%%%

\end{document}